\documentclass[twocolumn,floatfix,showpacs,prl]{revtex4}%
\usepackage{graphicx}%
\usepackage{amsmath}%
\usepackage{amsfonts}%
\usepackage{amssymb}
\usepackage{bm}
\usepackage{color}

\def\s{{\sigma}}
\def\e{{\epsilon}}
\def\k{{ {\bm k} }}
\def\p{{ {\bm p} }}
\def\q{{ {\bm q} }}
\def\Q{{ {\bm Q} }}

\def\0{{ {\bm 0} }}
\def\w{{\omega}}
\def\a{{\alpha}}
\def\b{{\beta}}

\def\dxz{{xz}}
\def\dyz{{yz}}
\def\dxy{{xy}}

\allowdisplaybreaks[4]

\begin{document}
\title{
Nematicity, magnetism and superconductivity in FeSe under pressure: \\
Unified explanation based on the self-consistent vertex correction theory
}
\author{
Youichi Yamakawa,
and Hiroshi Kontani
}

%\email[]{kon@slab.phys.nagoya-u.ac.jp}

\date{\today }

\begin{abstract}
To understand the
rich electronic phase diagram in FeSe under pressure
that vividly demonstrates the strong interplay between the 
nematicity, magnetism and superconductivity,
we analyze the electronic states by including the higher-order
many-body effects called the vertex correction (VC).
We predict the
pressure-induced emergence of $\dxy$-orbital hole-pocket
based on the first-principles analysis.
Due to this pressure-induced Lifshitz transition,
%Here, we construct the multiorbital Hubbard model for FeSe under pressure
%by referring to the first-principles calculations,
%and analyze the electronic states by including the higher-order
%many-body effects called the vertex correction (VC).
%When the pressure-induced $\dxy$-orbital Fermi pocket appears,
the spin fluctuations on the $\dxy$ orbital are enhanced,
whereas those on $\dxz,\dyz$ orbitals are gradually reduced.
For this reason, nonmagnetic orbital order $O=n_{xz}-n_{yz}$,
which is driven by the spin fluctuations on $\dxz,\dyz$ orbitals 
through the intra-orbital VCs, is suppressed, 
and it is replaced with the magnetism of $\dxy$-orbital $d$-electrons.
The nodal $s$-wave state at ambient pressure ($O\ne0$)
and the enhancement of $T_{\rm c}$ under pressure
are driven by the cooperation between the spin and orbital fluctuations.

\end{abstract}

\address{
Department of Physics, Nagoya University,
Furo-cho, Nagoya 464-8602, Japan. 
}
 

\sloppy

\maketitle

%%%%%%%%%%%%%%%%%%
%Introduction
%%%%%%%%%%%%%%%%%%
\section{Introduction}

Close relationship between the nematicity, magnetism,
and high-$T_{\rm c}$ superconductivity
is an essential electronic properties in Fe-based superconductors.
%The high-$T_{\rm c}$ state is realized near 
%the magnetic order and the electronic nematic order.
%The latter is the 
%spontaneous  
The nematic transition is the 
rotational symmetry breaking 
in the electronic states at $T_{\rm str}$,
and the strong nematic fluctuations are observed above $T_{\rm str}$
\cite{Yoshizawa,Egami,Gallais}. 
Inside the nematic phase below $T_{\rm str}$, 
large orbital polarization ($E_{yz}-E_{xz}\sim60$meV)
is observed by ARPES studies
%the angle-resolved-photoemission spectroscopy (ARPES)
\cite{Shen}.
%\cite{Shen,FeSe-ARPES2,FeSe-ARPES22,FeSe-ARPES3,FeSe-ARPES4,FeSe-ARPES5,FeSe-ARPES6}.
%Below $T_{\rm str}$, 
%the magnetic order appears except for Fe(Se,S)
%\cite{FeSe-ARPES2,FeSe-ARPES22,FeSe-ARPES3,FeSe-ARPES4,FeSe-ARPES5,FeSe-ARPES6}
As the nematic order parameter,
the spin-nematic order 
\cite{Kivelson,Fernandes}
and orbital/charge order
\cite{Kruger-OO,Lee-OO,Onari-SCVC}
have been discussed actively.
In both scenarios \cite{Kivelson,Fernandes,Onari-SCVC},
the nematicity and magnetism are closely related, so
%the nematicity is driven by the spin fluctuations, so
the magnetic transition slightly below $T_{\rm str}$
observed in many Fe-based superconductors
is naturally explained.
%the nematicity and magnetism are considered to be closely related.

%To achieve deep understanding 
To uncover the origin of the 
nematicity in Fe-based superconductors,
%to uncover the true nematic order parameter,
very rich phase diagram of FeSe attracts increasing attention.
Above $T_{\rm str}=90$K in FeSe at ambient pressure,
strong nematic fluctuations emerges whereas 
low-energy spin fluctuations are weak
\cite{Baek,Ishida-NMR,Raman-Gallais,Shibauchi-nematic}.
In the nematic state below $T_{\rm str}$,
large orbital polarization emerges
\cite{FeSe-ARPES2,FeSe-ARPES22,FeSe-ARPES3,FeSe-ARPES4,FeSe-ARPES5,FeSe-ARPES6},
whereas
no magnetism appears down to the superconducting temperature $T_{\rm c}=9$K.
%and very unusual ``sign-reversion orbital polarization'' is observed
%by ARPES studies.
Both the spin nematic 
\cite{Valenti-FeSe,Wang-nematic,Yu-nematic,Fanfarillo}
and the orbital order 
\cite{Yamakawa-FeSe,Onari-FeSe,Jiang-FeSe,Chubukov-nemaic2,Fanfarillo-FeSe}
scenarios have been applied for FeSe so far.
In the latter scenario,
the orbital nematicity is unable to be explained unless the 
higher-order electronic correlations beyond the mean-field,
called the vertex corrections (VCs),
is taken into consideration.
%by the mean-field-level theories such as the random-phase-approximation (RPA).
Recently, 
it was revealed that 
the Aslamazov-Larkin (AL) type VC,
which expresses the strong interference between the orbital and spin fluctuations,
gives rise to the nematicity due to the orbital order $O=n_{xz}-n_{yz}$.
The nontrivial ``sign-reversal orbital polarization''
observed by ARPES studies below $T_{\rm str}$ \cite{FeSe-ARPES6}
is reproduced by taking both the AL-VC and the Maki-Thompson VC
\cite{Onari-FeSe}.

The pressure-induced change in the electronic states in FeSe 
is an important open problem
\cite{Imai,Takano-rev,mSR1,mSR2,Terashima-pFeSe,Bohmer-pFeSe,NMR-pFeSe,Shibauchi-pFeSe}.
%attracts increasing attention.
%At ambient pressure, $T_{\rm str}=90$K and 
%the superconductivity occurs at $T_c=9$K.
Under pressure, $T_{\rm str}$ decreases whereas $T_c$ increases,
%By applying the pressure, $T_{\rm str}$ decreases from 90K 
%30K at $P\sim2$GPa.
and the nematic phase is replaced with the magnetic order ($T_m\sim30$K)
at $P=2$GPa.
This pressure-induced magnetism in FeSe may remind us of the
second magnetic phase in heavy H-doped region in LaFeAsO \cite{Onari-SCVCS}.
Around 4GPa, $T_m$ reaches its maximum ($45$K),
and $T_c$ in the magnetic phase exceeds $20$K.
It is a significant challenge for theorists to explain such 
rich pressure-induced electronic states based on the 
realistic Hubbard model for FeSe.

In this paper, 
%we study the origin of the nematicity, magnetism, and superconductivity
%in FeSe based on the self-consistent VC (SC-VC) theory
%\cite{Onari-SCVC}.
we predict the
pressure-induced emergence of $\dxy$-orbital hole-pocket
based on the first-principles analysis.
Due to this pressure-induced Lifshitz transition,
the spin fluctuations on the $\dxy$ orbital increase,
whereas those on the $\dxz,\dyz$ orbitals gently decrease.
%according to the self-consistent VC (SC-VC) theory
\cite{Onari-SCVC}.
For this reason, nematic orbital order $n_{xz}\ne n_{yz}$,
which is driven by the spin fluctuations on $\dxz,\dyz$ orbitals 
through the intra-orbital VCs,
%orbital-spin interplay due to the AL-VC,
is suppressed and replaced with the magnetism on the $\dxy$ orbital.
The nodal $s$-wave state at 0GPa below $T_{\rm str}$
\cite{Song,Hanaguri,Feng-FeSe-gap}
and the enhancement of $T_{\rm c}$ under high pressures
are given by the orbital+spin fluctuations.
The close interplay between the 
nematicity, magnetism, and superconductivity due to the VC in FeSe
would be the universal feature of Fe-based superconductors.

In FeSe at ambient pressure, 
weak spin fluctuations are enough for 
realizing the orbital order due to the AL-VC,
because of the smallness of the ratio $J/U$
and the ``absence of the $\dxy$-orbital hole-pocket''
as discussed in Ref. \cite{Yamakawa-FeSe}.
Since the Fermi energy of each pocket is very small in FeSe,
the nature of spin fluctuations is
sensitively modified by the pressure-induced parameter change.
Then, spin-fluctuation-driven orbital order is also sensitively modified.
Therefore, rich $T$-$P$ phase diagram in FeSe is naturally explained.

%The rich T-P phase diagram in FeSe manifests the significance of the
%interplay between orbital and spin fluctuations 
%in Fe-based superconductors.

We study the following two-dimensional eight-orbital Hubbard model:
\begin{eqnarray}
H= H_0 +H_P+ r H_U
\label{eqn:Ham} ,
\end{eqnarray}
where $H_0$ is the kinetic term for FeSe at ambient pressure
introduced in Ref. \cite{Yamakawa-FeSe},
and $H_U$ is the first-principles multiorbital interaction for FeSe
\cite{Arita}.
In FeSe, ${\bar U}=7.21$eV and ${\bar J}=0.68$eV on average.
Here, $r$ is the reduction factor for the interaction term
\cite{Yamakawa-FeSe}.
We denote 
$d_{z^2},d_{xz},d_{yz},d_{xy},d_{x^2-y^2}$ orbitals as 1, 2, 3, 4, 5,
and $p_x,p_y,p_z$ orbitals as 6, 7, 8.
We introduce the renormalization factor 
%$d_{xy}-orbital$
$z_{4}=1/1.2$ and $z_{l}=1$ for $l\ne4$
\cite{Yamakawa-FeSe},
consistently with the microscopic theory 
in Ref. \cite{DMFT1}.

In Eq. (\ref{eqn:Ham}), 
$H_P$ represents the change in the kinetic term under pressure.
To derive $H_P$,
we perform the band calculation using the crystal structure of 
FeSe under pressure
\cite{Millican,Bohmer}:
The most dominant pressure-induced change 
is the lift of the $\dxy$-orbital level around $\q=(\pi,\pi)$
due to the increase of the Se-atom height \cite{Kuroki-z,Haule}.
To shift the $E_{xy}$-level by $\Delta E_{xy}$ at $\q=(\pi,\pi)$,
we set $H_P = \sum_{i,j,\s} \delta t_{i,j}c^\dagger_{i,4\s}c_{j,4\s}$,
where $\delta t_{i,j}=\Delta E_{xy}/4$ for the on-site ($i=j$),
$\delta t_{i,j}=-\Delta E_{xy}/8$ and $\delta t_{i,j}=\Delta E_{xy}/16$ 
for the first- and second-nearest sites \cite{Yamakawa-FeSe};
see the Supplemental Material (SM):A \cite{SM}.
Figure \ref{fig:fig1} (a) shows the bandstructure for 
$\Delta E_{xy}=0$, 0.06, and 0.12 eV, given by the solutions of 
${\rm det}\{ {\hat Z}\cdot\e-({\hat H}_0(\k)+{\hat H}_P(\k))\}=0$,
or equivalently given by the eigenvalues of
${\hat Z}^{-1/2}({\hat H}_0(\k)+{\hat H}_P(\k)){\hat Z}^{-1/2}$,
where ${Z}_{l,m}=(1/z_l)\delta_{l,m}$ \cite{Yamakawa-FeSe,Onari-FeSe}.

The Fermi surfaces (FSs) for $\Delta E_{xy}=0$ and 0.12 eV
are shown in Figs. \ref{fig:fig1} (b) and (c), respectively.
In the present model, $\dxy$-orbital pocket appear 
for $\Delta E_{xy}\gtrsim 0.1$eV.
According to the first principles study,
$\Delta E_{xy}\sim0.1$eV is realized under $\sim4$ GPa \cite{SM}.
In FeSe at ambient pressure, 
the top of the $\dxy$-hole pocket is about $40$meV below the Fermi level
experimentally \cite{FeSe-ARPES6},
which corresponds to $\Delta E_{xy}\sim 0.05$eV in the present model.

%%%%%%%%%%%%%%%%%%%%%%%%%%%%%%%%%
\begin{figure}[!htb]
\includegraphics[width=.7\linewidth]{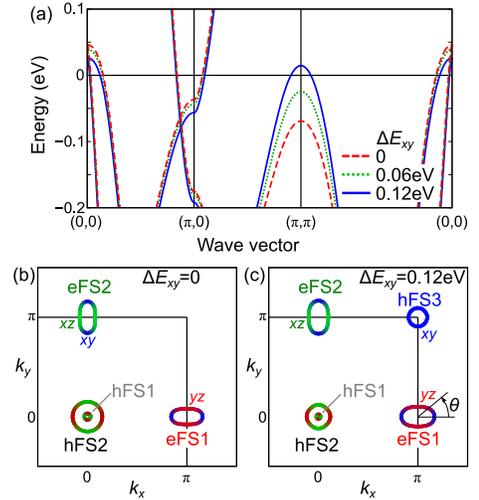}
\caption{
(color online)
(a) Band structure of the present FeSe model for 
$\Delta E_{xy}=0$, 0.06 and and 0.12 eV.
(b) FSs for $\Delta E_{xy}=0$ and 
(c) FSs for $\Delta E_{xy}=0.12$eV.
The hFS3 is the $\dxy$-orbital hole-pocket.
The colors green, red, and blue represents the 
$\dxz$, $\dyz$, and $\dxy$ orbitals, respectively.
}
\label{fig:fig1}
\end{figure}
%%%%%%%%%%%%%%%%%%%%%%%%%%%%%%%%%

Hereafter, we analyze the model Hamiltonian (\ref{eqn:Ham})
based on the self-consistent VC (SC-VC) theory.
The spin- or charge-channel susceptibility is given as
\begin{eqnarray}
{\hat \chi}^x(q)
={\hat \Phi}^x(q)[{\hat 1}-{\hat U}^{0x}{\hat \Phi}^x(q)]^{-1}
 \ \ \ (x=s,c)
\label{eqn:chi} ,
\end{eqnarray}
where $x=s$ or $c$, and $q=(\w_l=2\pi l T, \q)$.
${\hat U}^{0x}$ is the first-principles Coulomb interaction for FeSe
derived in Ref. \cite{Arita}, and it was introduced as ${\hat \Gamma}^x$
in Ref. \cite{Yamakawa-FeSe}.
%see the SM \cite{SM}.
The irreducible susceptibility is
${\hat \Phi}^x(q)= {\hat \chi}^0(q)+{\hat X}^x(q)$,
where ${\hat \chi}^0(q)$ is the bare susceptibility and 
${\hat X}^x(q)$ is the VC that is dropped in the 
random-phase-approximation (RPA).
The spin (charge) Stoner factor $\a_{S(C)}$ is defined as the 
largest eigenvalue of ${\hat U}^{0s(c)}{\hat \Phi}^{s(c)}(\q)$:
The magnetic order (orbital order) is established 
when $\a_{S(C)}=1$.

\begin{figure}[!htb]
\includegraphics[width=.99\linewidth]{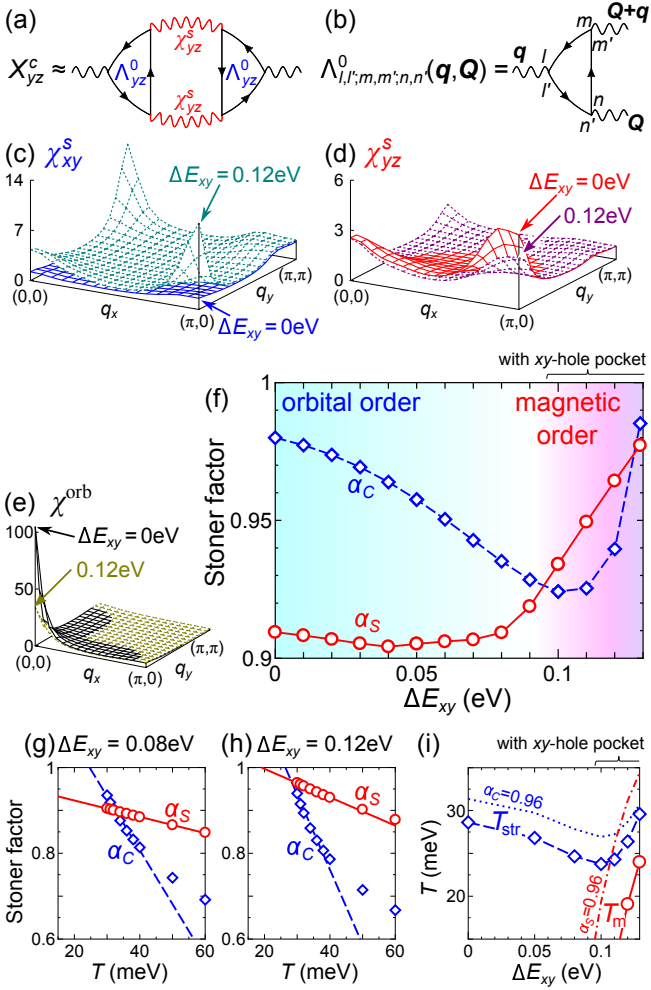}
\caption{
(color online)
(a) Charge-channel AL-type VC for the $\dyz$-orbital and 
(b) three-point vertex.
(c) $\chi^s_{xy}(\q)$,
(d) $\chi^s_{yz}(\q)$, and 
(e) $\chi^{\rm orb}(\q)$
or $\Delta E_{xy}=0$ and $0.12$eV at $r=0.266$ and $T=30$meV.
(f) Stoner factors as functions of $\Delta E_{xy}$
at $r=0.266$ and $T=30$meV.
(g) $T$-dependences of $\a_{S,C}$ for $\Delta E_{xy}=0.08$ eV and 
(h) those for $\Delta E_{xy}=0.12$ eV.
(i) $T_m$ and $T_{\rm str}$ defined by the relation $\a_{S,C}=1$ and $0.96$.
%Weiss temperatures $\theta_S$ and $\theta_C$.
}
\label{fig:fig2}
\end{figure}

In the SC-VC theory, we calculate both the Maki-Thompson-type VC
and the AL-type VC, which are the first-order and the 
second-order diagrams with respect to $\chi^{x}$, respectively
\cite{Onari-SCVC,Yamakawa-FeSe}.
%In Fe-based superconductors,
Figure  \ref{fig:fig2} (a) shows the diagrammatic expression 
for the AL-type VC $X^c_{yz}({\bm 0})\sim 
|\Lambda^0_{yz}|^2T\sum_{\q}\{\chi^s_{yz}(\q)\}^2$.
Here, ${\hat \Lambda}^0_{yz}$ is the three-point vertex for $yz$-orbital,
given by ${\hat \Lambda}^0_{3,3;3,3;3,3}({\bm 0};\Q)$ 
in Fig. \ref{fig:fig2} (b)
\cite{Onari-SCVC}.
Thus, $X^c_{yz}({\bm 0})$ increases in proportion to 
$\sum_{\q}\{\chi^s_{yz}(\q)\}^2\sim \chi^s_{yz}(\Q)$.
%in the presence of spin fluctuations on $\dxz,\dyz$-orbitals.
The increment of the AL-VC drives the
strong nematic orbital susceptibility for $O=n_{xz}-n_{yz}$, 
$\chi^{\rm orb}({\bm 0}) \equiv \sum_{l,m}^{2,3}(-1)^{l+m}\chi^c_{l,l;m,m}({\bm 0})$.
% when spin fluctuations on $\dxz,\dyz$-orbitals.
%Here, we drop ${\hat X}^s(q)$ 
Here, we calculate only the charge-channel VC $X_{l,l';m,m'}^c(q)$
for $l,l',m,m'=2\sim4$ except for $l=l'=m=m'=4$,
by following Ref. \cite{Onari-SCVC}, which is justified
%for analyzing the orbital fluctuations 
for various Fe-based superconductors
%as we verified in our previous studies
\cite{Onari-SCVC,Yamakawa-FeSe}.
In the SM:B \cite{SM}, we explain that 
essentially similar results are obtained if we perform 
the self-consistent calculation for 
both ${\hat X}^{s}(q)$ and ${\hat X}^c(q)$ for all $d$-orbitals.

%%%%%%%%%%%%%%%%%%%%%%%%%%%%%%%%%
%%%%%%%%%%%%%%%%%%%%%%%%%%%%%%%%%

In Fig. \ref{fig:fig2} (c), we show the 
spin susceptibility on $xy$-orbital,
$\chi^s_{xy}(\q) \equiv \chi^s_{4,4;4,4}(\q)$,
%and $\chi^s_{xy}(\q) \equiv \chi^s_{4,4;4,4}(\q)$, respectively,
at $r=0.266$ and $T=30$meV.
%and the orbital susceptibility for $O=n_{xz}-n_{yz}$,
%$\chi^{\rm orb}(\q) \equiv \sum_{l,m}^{2,3}(-1)^{l+m}\chi^c_{l,l;m,m}(\q)$.
%We set $\Delta E_{xy}=0$ in (b), and 
%$\Delta E_{xy}=0.12$eV in (c), for $r=0.266$.
For $\Delta E_{xy}=0.12$eV, the hFS3 at $(\pi,\pi)$ appears,
and therefore $\chi^s_{xy}(\Q)$ $(\Q=(\pi,0))$ is drastically enlarged
due to the good nesting between eFS1,2 and the hFS3
on $\dxy$-orbital.
We stress that the emergence of the hFS3
could lead to the inhomogeneous phase separation
as discussed in Ref. \cite{Nori}.
In contrast, 
$\chi^s_{yz}(\q) \equiv \chi^s_{3,3;3,3}(\q)$ 
%$\chi_{yz}(\Q)$
%spin susceptibility on $yz$-orbital 
in Fig. \ref{fig:fig2} (d) gently decreases
with $\Delta E_{xy}$, since the nesting condition
on $\dyz$-orbital becomes worse due to the 
shrinkage (expansion) of hFS1,2 (eFS1,2).
This suppression of $\chi_{yz}(\Q)$ caused the 
reduction of the orbital susceptibility in Fig. \ref{fig:fig2} (e),
due the reduction of the AL-term 
$X_{yz}^c({\bm 0})\propto \chi^s_{yz}(\Q)$.
Such drastic $\Delta E_{xy}$-dependences of the 
spin and orbital susceptibilities 
are contributed by the
the smallness of the FSs in FeSe.

%For $\Delta E_{xy}=0$,
%the strengths of $\chi^s_{yz}$ and $\chi^s_{xy}$ are comparable.
%In this case, moderate spin fluctuations on $\dxz,\dyz$-orbitals
%gives rise to the large $\chi^{\rm orb}({\bm 0})$.
%For $\Delta E_{xy}=0.12$eV,
%the spin fluctuations develop only on the $\dxy$-orbital 
%[$\chi^s_{xy}\gg\chi^s_{xz(yz)}$] because of the good nesting on $\dxy$-orbital
%between eFS1,2 and the hFS3.
%As a result,
%$\chi^{\rm orb}({\bm 0})$ is reduced to be comparable to $\chi^s_{xy}(\pi,0)$.

Figure \ref{fig:fig2} (f) shows the Stoner factors 
$\a_{S,C}$ obtained by the SC-VC theory for $r=0.266$ at $T=30$meV. 
The obtained $\a_{S(C)}$ is qualitatively proportional to $T_{m({\rm str})}$.
At $\Delta E_{xy}=0$, which corresponds to $P=0$GPa,
$\a_C$ is close to the unity whereas $\a_S\sim0.9$.
%For $\Delta E_{xy}<0.1$eV, 
%$\a_C$ decreases in proportion to $\Delta E_{xy}$, 
%reflecting the decrease of $X_{yz}({\bm 0}]])\sim \chi^s_{yz}(\Q)$.
%consistently with the experimental reduction of $T_{\rm str}$ 
%under $P=0\sim2$GPa.
%add
When $\Delta E_{xy}>0$, the chemical potential $\mu$ increases 
even for $\Delta E_{xy}\ll0.1$ eV due to the finite-$T$ effect.
Then, the nesting between hFS1,2 and eFS1 becomes worse, 
and therefore $\chi^s_{yz}(\Q)$ gradually decreases.
At the same time, $\a_C$ decreases since 
$X^c_{yz}({\bm 0})\sim 
|\Lambda^0_{yz}|^2T\sum_{\q}\{\chi^s_{yz}(\q)\}^2$ becomes smaller,
%the AL-term on $yz$-orbital becomes smaller, 
%as shown in Fig. \ref{fig:fig2} (f).
consistently with the experimental reduction of $T_{\rm str}$ under $P=0\sim2$GPa.
In contrast, $\chi^s_{xy}(\Q)$ starts to increase 
for $\Delta E_{xy}\gtrsim 0.08$ eV, due to the nesting 
between the hFS3 and eFS1,2 on $xy$-orbital.
For this reason, $\a_S$ rapidly increases in Fig. \ref{fig:fig2} (f).
When $\Delta E_{xy}\gtrsim0.1$ eV, $\a_C$ starts to increase since 
$\chi^s_{xy}(\q)$ also contribute to the ``AL-term on $yz$-orbital
through $|\Lambda_{yz\mbox{-}xy}^0|^2T\sum_{q}{\chi^s_{xy}(q)}^2$'',
where $\Lambda_{yz\mbox{-}xy}^0\equiv \Lambda_{3,3;4,4;4,4}^0({\bm0};\Q)$.
Although the off-diagonal three-point vertex $\Lambda_{yz\mbox{-}xy}^0$
is much smaller than the diagonal one $\Lambda_{yz}^0$, 
it is finite since the electron-FSs are composed of 
three orbitals $2 \sim 4$.

%For $\Delta E_{xy}>0.1$eV, 
%$\a_S$ starts to increase due to the $\dxy$-orbital nesting
%between the pressure-induced hFS3 and eFS1,2.
%This result is consistent with the 
%increment of $1/T_1T$ under pressure \cite{Imai}
%and the emergence of magnetism for $P>2$GPa 
%\cite{mSR1,mSR2,Terashima-pFeSe,Bohmer-pFeSe,NMR-pFeSe,Shibauchi-pFeSe}.
%In the region $\a_S>\a_C$, the tetragonal ($C_4$) magnetic phase 
%may appear for $T_m>T>T_{\rm str}$ if the spin-lattice coupling is small.
%Interestingly, $\a_C$ also starts to increase for $\Delta E_{xy}>0.1$eV
%since $\chi^s_{xy}(\q)$ also contribute to the ``AL-term on $yz$-orbital''
%due to $|\Lambda_{3,3;4,4;4,4}^0|^2T\sum_{\q}\{\chi^s_{xy}(\q)\}^2$.
%Here, the off-diagonal three-point vertex $\Lambda_{3,3;4,4;4,4}^0$ 
%is small but finite 
%since the electron-FSs are composed of $2\sim4$ orbitals.

To confirm the validity of the simplified pressure term $H_P$
introduced in the present study, 
we derive the pressure term 
from the first principles study, $H_P^{\rm 1st}$, 
and perform the numerical study based on the SC-VC theory
in the SM: D \cite{SM}.
The obtained $P$-dependences of $\a_S$ and $\a_C$
are consistent with Fig. \ref{fig:fig2} (f).

We also estimate the transition temperatures $T_m$ and $T_{\rm str}$.
As shown in Figs. \ref{fig:fig2} (g) and (h),
the Stoner factors $\a_{S(C)}$ follow the $T$-linear relations
for $T\gtrsim30$meV,
at $\Delta E_{xy}=0.08$ eV and $0.12$ eV.
(Then, $\chi^s(\Q)$ and $\chi^{\rm orb}({\bm 0})$ 
follow the Curie-Weiss relations.)
By extrapolating the $T$-linear relations,
we estimate $T_m$ and $T_{\rm str}$
under the condition $\a_{S,C}=1$ and $\a_{S,C}=0.96$
in Fig. \ref{fig:fig2} (i).
In both cases, 
$T_{\rm str}$ decreases with $\Delta E_{xy}$ for $\Delta E_{xy}<0.1$ eV,
whereas $T_m$ drastically increases for $\Delta E_{xy}>0.1$ eV,
consistently with experimental phase diagram.
In the region $T_m>T>T_{\rm str}$ for $\a_{S,C}=0.96$ with $\Delta E_{xy}>0.11$ eV,
the tetragonal ($C_4$) stripe magnetic phase appears,
if the spin-lattice coupling is negligibly small.
%Even if the spin-lattice coupling is large,
%the $C_4$ magnetic/charge order is possible as discussed in Refs.
% \cite{C4-1,C4-2}.

%This enhancement of $\a_C$ above $\Delta E_{xy}\approx0.1$eV
%is robust against model parameters.
%In this model, $\a_{S,C}$ diverges for $\Delta E_{xy}>0.13$eV,
%which means that $T_m$ or $T_{\rm str}$ is higher than $T=30$meV
%under high pressure.
%However, the transition temperature will be lowered 
%if we consider the increment of the bandwidth,
%which is a minor pressure effect and is neglected in this study 
%for simplicity.
%This is our future issue.

%Figure \ref{fig:fig2} (e) shows the Weiss temperatures $\theta_{S,C)$
%given by the Curie-Weiss fitting of the enhancement factors 
%$(1-\a_{S,C})^{-1} \propto (T-\theta_{S,C))^{-1}$
%for $T=30\sim50$ meV.
%Here, $\theta_{S(C))$ gives an indication of $T_{m(S)}$.

%This enhancement of $\a_C$ above $\Delta E_{xy}\approx0.1$eV
%is robust against model parameters.
%In this model, $\a_{S,C}$ diverges for $\Delta E_{xy}>0.13$eV,
%which means that $T_m$ or $T_{\rm str}$ is higher than $T=30$meV
%under high pressure.
%However, the transition temperature will be lowered 
%if we consider the increment of the bandwidth,
%which is a minor pressure effect and is neglected in this study 
%for simplicity.
%This is our future issue.
%In Fig. \ref{fig:fig2} (d),
%$\lambda_{\rm tot}$ is the eigenvalue of the gap equation;
%see below.

%%%%%%%%%%%%%%%%%%%%%%%%%%%%%%%%%
\begin{figure}[!htb]
\includegraphics[width=.85\linewidth]{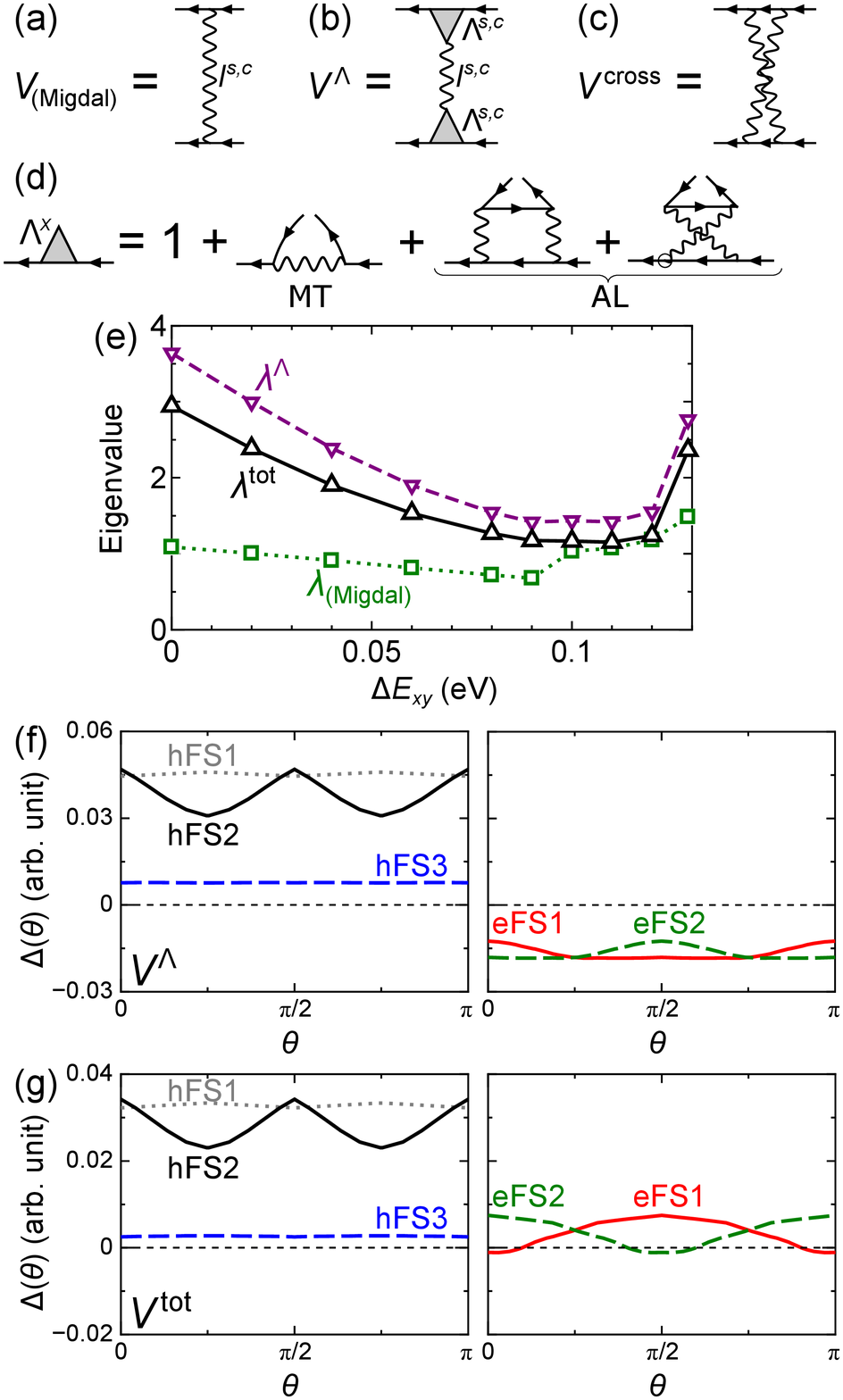}
\caption{
(color online)
%(a) Gap equation for the total pairing interaction $V^{\rm tot}$.
(a) $V_{\rm (Migdal)}$, (b) $V^{\Lambda}$, and (c) $V^{\rm cross}$.
(d) $U$-VC $\Lambda^x$.
(e) Eigenvalues for various pairing interactions:
$\lambda^{\rm tot}$ for $V^{\rm tot}$,
$\lambda^{\Lambda}$ for $V^{\Lambda}$, and
$\lambda_{(\rm Migdal)}$ for $V_{(\rm Migdal)}$.
(f) $s_\pm$-wave gap given by $V^{\Lambda}$
for $\Delta E_{xy}=0.12$eV.
(g) $s_{++}$-wave gap given by $V^{\rm tot}$.
The FSs are shown in Fig. \ref{fig:fig1} (c).
%Nonmagnetic state is assumed in the numerical study.
Numerical study is performed in the nonmagnetic state.
}
\label{fig:fig3}
\end{figure}
%%%%%%%%%%%%%%%%%%%%%%%%%%%%%%%%%

In the next stage, we study the superconducting state.
Both the spin 
\cite{Kuroki-z,Mazin,Chubukov,Graser,Chubu-Rev} 
and/or orbital 
\cite{Kontani-RPA,Onari-SCVCS,Yamakawa-eFeSe}
fluctuation mediated mechanisms have been studied.
%Here, we analyze the following ``beyond-Migdal-Eliashberg gap equation'' 
%introduced in Ref. \cite{Yamakawa-eFeSe}:
The linearized gap equation is
\begin{eqnarray}
\lambda\Delta_\a(k)=-T\sum_{p,\b}V_{\a,\b}^{\rm SC}(k,p)|G_\b(p)|^2\Delta_\b(p)
\label{eqn:gapeq} ,
\end{eqnarray}
where $k=((2n+1)\pi T,\k)$,
$\Delta_\a(k)$ is the gap function on the $\a$-FS,
and $\lambda$ is the eigenvalue that reaches unity at $T=T_c$.
$V_{\a,\b}^{\rm SC}$ is the pairing interaction.
In the conventional Migdal approximation,
${\hat V}^{\rm SC}$ in the orbital basis is given as
${\hat V}_{(\rm Migdal)}(k,p)=\frac32{\hat I}^s(k-p)
+\frac12{\hat I}^c(k-p)-{\hat U}^{0s}$,
where 
${\hat I}^{x}(q)= {\hat U}^{0x}+
{\hat U}^{0x}{\hat \chi}^{x}(q){\hat U}^{0x}$ $(x=s,c)$.
%using the matrix $u_{l,\a}(\k)=\langle l,\k|\a,\k\rangle$.
It is schematically shown in Fig. \ref{fig:fig3} (a).
However, since this approximation is not justified for 
Fe-based superconductors,
we introduce the following ``beyond-Migdal-Eliashberg pairing interaction'' 
introduced in Ref. \cite{Yamakawa-eFeSe}:
\begin{eqnarray}
&&{\hat V}^{\rm tot}(k,p)={\hat V}^{\Lambda}(k,p)+
{\hat V}^{\rm cross}(k,p)
\label{eqn:V-total} ,
\end{eqnarray}
which are depicted in Figs. \ref{fig:fig3} (b) and (c).
${\hat \Lambda}^x$ in Fig. \ref{fig:fig3} (b) 
is the VC for the coupling constant ${\hat U}^{0x}$,
called the $U$-VC.
Since $|\Lambda^c|^2\gg1$ due to the AL-type VC
shown in Fig. \ref{fig:fig3} (d)
\cite{Yamakawa-eFeSe,Tazai},
moderate orbital fluctuations give sizable attractive interaction.
Also, $V^{\rm cross}$ is the ``AL-type crossing fluctuation interaction''
%which gives large attractive interaction
%in Fe-based superconductors 
\cite{Yamakawa-eFeSe}.
The analytic expressions for Figs. \ref{fig:fig3} (b)-(d)
are explained in Refs. \cite{Tazai,Yamakawa-eFeSe}
and in the SM: E \cite{SM}.
%he significance of the $U$-VC in $V^\Lambda$ has been verified 
%in various multiorbital Hubbard models 
%\cite{Onari-SCVC,Yamakawa-eFeSe,Tazai}.
In Ref. \cite{Tazai}, we verified the significance of the $U$-VC 
by applying the functional-renormalization-group (fRG) method to 
the two-orbital Hubbard model.
The fRG method enables us to generate the higher-order VCs
(including the higher-order processes other than Figs. \ref{fig:fig3} (a)-(d))
in a systematic and unbiased way.
%We showed that both the ``k-dependence'' and the 
%``spin/charge-channel dependence'' of the pairing interaction given by 
%the fRG four-point vertex are well approximated by the 
%``single-fluctuation-exchange approximation with the U-VCs''. 
%Although $V^{cross}$ gives large attractive interaction in 
%electron-doped FeSe model, its importance seems to be model-dependent. 

%In addition, we also introduce the interaction
%%
%\begin{eqnarray}
%{\hat V}^{\rm no-cross}(k,p)={\hat V}^{\rm tot}(k,p)-V^{\rm cross}(k,p)
%\label{eqn:V-nocross}
%\end{eqnarray}
%%
%to clarify the importance of the crossing term.

%The $s_\pm$-wave with sign reversal between electron- and hole-pockets
%is obtained for both $V_{(\rm Migdal)}$ and $V^{\Lambda}$.
%The gap function for $V^{\Lambda}$ at $\Delta E_{xy}=0.12$eV
%is shown in Fig. \ref{fig:fig3} (c).

The eigenvalue $\lambda^{\rm tot}$ obtained for $V^{\rm tot}$
in the absence of the nematicity and magnetism 
is shown in Fig. \ref{fig:fig3} (e),
together with $\lambda_{(\rm Migdal)}$ for $V_{(\rm Migdal)}$ and
$\lambda^{\Lambda}$ for $V^{\Lambda}$:
The relation $\lambda_{(\rm Migdal)} \ll \lambda^\Lambda$
means that the charge $U$-VC strongly enlarges $T_{\rm c}$.
For $V^{\rm SC}=V^{\Lambda}$ at $\Delta E_{xy}=0.12$eV,
the $s_\pm$-wave state with sign reversal between electron- and hole-pockets
is obtained as shown in Fig. \ref{fig:fig3} (f).
In contrast, the ``$s_{++}$-wave state without sign reversal'' is obtained 
for the total pairing interaction $V^{\rm tot}$,
since $V^{\rm cross}$ gives large inter-pocket attractive interaction:
The obtained nodal $s_{++}$-wave state at $\Delta E_{xy}=0.12$eV
is shown in Fig. \ref{fig:fig3} (g).
The $s_{++}$-state will be stabilized by the 
pressure-induced strong $e$-ph interaction 
\cite{Haule}.

%%%%%%%%%%%%%%%%%%%%%%%%%%%%%%%%%
\begin{figure}[!htb]
\includegraphics[width=.85\linewidth]{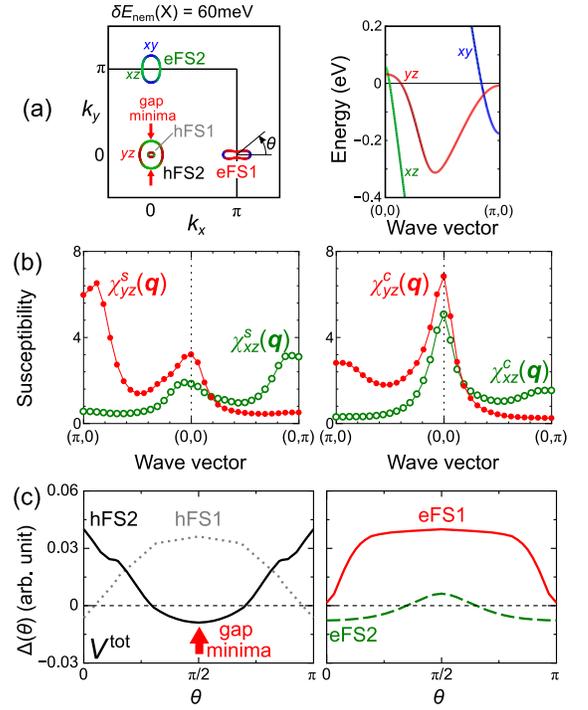}
\caption{
(color online)
(a) FSs and bandstructure in the presence of the 
sign-reversing orbital polarization
$\delta E_{\rm nem}(\Gamma)= -30$meV and $\delta E_{\rm nem}({\rm X})= 60$meV.
Here, $\Delta E_{xy}=0$.
(b) $\chi^s_{xz}(\q)$, $\chi^s_{yz}(\q)$, 
$\chi^c_{xz}(\q)$, and $\chi^c_{yz}(\q)$
obtained by the SC-VC theory.
%(c) $s$-wave gap functions on the FSs obtained for $V^{\Lambda}$.
(c) $s$-wave gap functions obtained for $V^{\rm tot}$:
The gap functions on each FS is nodal.
}
\label{fig:fig4}
\end{figure}
%%%%%%%%%%%%%%%%%%%%%%%%%%%%%%%%%

Finally, we study the superconductivity
in FeSe in the nematic state at $P=0$  \cite{Mukherjee}.
% ($\delta n=0$).
Figure \ref{fig:fig4} (a) shows the FSs and bandstructure
under the sign-reversing
orbital polarization reported in Ref. \cite{FeSe-ARPES6}:
We set  $\delta E_{\rm nem}\equiv E_{yz}-E_{xz} = -30$meV at $\Gamma$ point,
and $\delta E_{\rm nem}=60$meV at X and Y points
by following Ref. \cite{Yamakawa-FeSe}.
The susceptibilities obtained by the SC-VC theory
for $z_4^{-1}=1.2$ and $r=0.266$ at $T=30$ meV 
[$(\a_S,\a_C)=(0.914,0.934)$]
are shown in Fig. \ref{fig:fig4} (b).
By introducing $\delta E_{\rm nem}$,
$\chi^s_{yz}(\pi,0)$ increases whereas $\chi^s_{xz}(0,\pi)$ decreases 
as explained in Ref. \cite{Kontani-RPA}.
Then, the AL-term $X_{yz}^c(\q)$ is enlarged 
due to the large $\chi^s_{yz}(\pi,0)$.
%strong spin fluctuations on the $d_{yz}$-orbital.
The obtained strong ferro- and antiferro-orbital fluctuations 
give significant pairing interaction on the $\dyz$-orbital.

Figure \ref{fig:fig4} (c) shows the $s$-wave gap function 
obtained for $V^{\rm tot}$.
%The gap functions on the hole-FS and electron-FSs are essentially same,
Here, the gap is large on the FSs with large $\dyz$-orbital component,
since strong intra-pocket force and inter-pocket attractive one are
caused by large $\chi^c_{yz}(\q)$ at $\q\sim{\bm 0}$ and $\q\sim(\pi,0)$
in addition to $V^{\rm cross}$.
%The nodal gap structure reflects the smallness of 
%the fluctuations on $d_{xz},d_{xy}$-orbitals.
The position of the nodal part on hole-FS is consistent with 
the experimental report \cite{Feng-FeSe-gap}.
The nodal structure on $\Delta_{\rm hFS}$ is unchanged when
two Dirac cones appear near X-point for larger $\delta E_{\rm nem}({\rm X})$,
as we show in SM: C \cite{SM}.
Thus, the nodal gap structure is robust against model parameters,
whereas the ratio between $|\Delta_{\rm hFS}|$ and $|\Delta_{\rm eFS}|$
is sensitive to parameters,
% note that the anisotropy of the gap functions is 
%very sensitive to the relative strengths of spin and orbital fluctuations,
so it is our future problem to explain the experimental 
data quantitatively.

In summary,
we studied the electronic states in FeSe at ambient pressure and under pressure.
%We constructed the realistic Hubbard model for FeSe under pressure
%and analyzed the model using the self-consistent VC (SC-VC) theory.
We predicted that the hFS3 appears under pressure.
Due to this pressure-induced Lifshitz transition,
the spin fluctuations on the $\dxy$ orbital are enlarged,
whereas those on the $\dxz,\dyz$ orbitals are reduced gently.
Since the orbital order is
driven by the spin fluctuations on $\dxz,\dyz$ orbitals via the 
intra-orbital VCs,
the nematicity is suppressed and replaced with the magnetism on 
the $\dxy$ orbital electrons under pressure.
Also, both the nodal $s$-wave state in the nematic state at ambient pressure
and the enlargement of $T_{\rm c}$ under pressure
are satisfactorily explained, due to the
novel cooperation between spin and orbital fluctuations.

\acknowledgments

We are grateful to A. Chubukov, P.J. Hirschfeld, R. Fernandes,
J. Schmalian, D.L. Feng, S. Shin, Y. Matsuda, T. Shibauchi, 
K. Okazaki and T. Shimojima for useful discussions.
This study has been supported by Grants-in-Aid for Scientific 
Research from MEXT of Japan.
%Part of numerical calculations were
%performed on the Yukawa Institute Computer Facility.

%%%%%%%%%%%%%%%%%%%%%%%%
%references
%%%%%%%%%%%%%%%%%%%%%%%%

%\end{document}

%%%%%%%%%%%%%%%%%%%%%%%%%%%%%%%%%%%%%%%%%%%%%%%%%%%%%
\clearpage

\makeatletter
\renewcommand{\thefigure}{S\arabic{figure}}
\renewcommand{\theequation}{S\arabic{equation}}
\makeatother
\setcounter{figure}{0}
\setcounter{equation}{0}
\setcounter{page}{1}
\setcounter{section}{1}

\begin{widetext}
\begin{center}
{\Large [Supplementary Material]}
\end{center} 

\begin{center}
{\large
\textbf{
Nematicity, magnetism and superconductivity in FeSe under pressure: \\
Unified explanation based on the self-consistent vertex correction theory
}}
\end{center} 
\begin{center}
Youichi Yamakawa, 
Hiroshi Kontani
\end{center} 

%\begin{center}
%\textit{Department of Physics, Nagoya University, Nagoya 464-8602, Japan}
%\end{center} 

\end{widetext}

%%%%%%%%%%%%%%%%%%%%%%%%%%%%%%%%%%%%%%%%%%%%%%%%%%%%%%%%
\subsection{A: First principles model Hamiltonian}
% Band calculation for FeSe under pressure}

In the main text, we studied the electronic states in FeSe under pressure
by using the SC-VC theory.
To describe the change in the bandstructure under pressure,
we introduced the additional term $H_P$
into the model Hamiltonian for FeSe at ambient pressure.
We set 
$H_P = \sum_{i,j,\s} \delta t_{i,j}c^\dagger_{i,4\s}c_{j,4\s}$,
where $\delta t_{i,j}=\Delta E_{xy}/4$ for the on-site ($i=j$),
$\delta t_{i,j}=-\Delta E_{xy}/8$ and $\delta t_{i,j}=\Delta E_{xy}/16$ 
for the first- and second-nearest sites.
By introducing $H_P$, the $d_{xy}$-orbital level around $(\pi,\pi)$
is lifted by $\Delta E_{xy}$.

\begin{figure}[!htb]
\includegraphics[width=.9\linewidth]{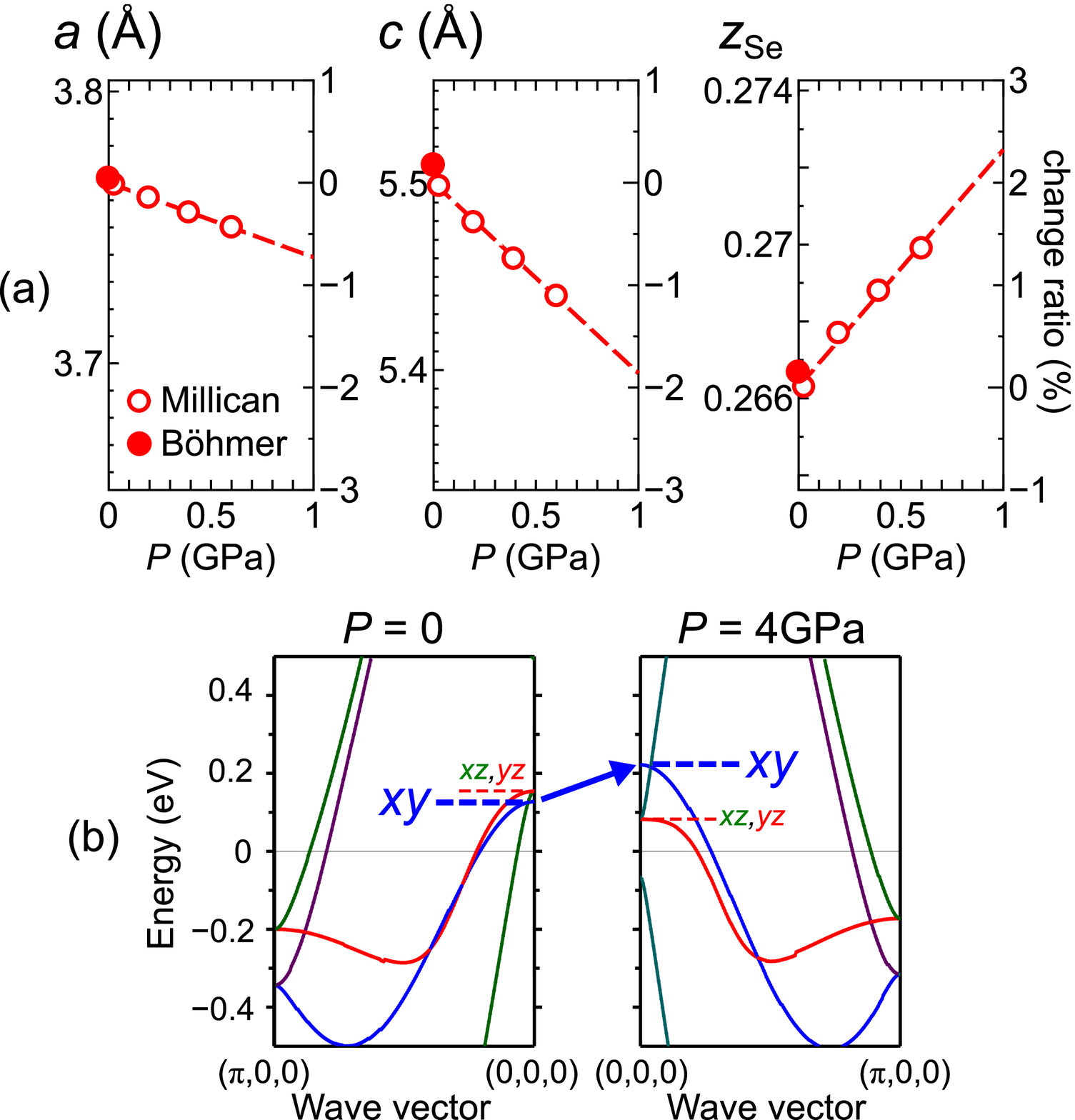}
\caption{
(color online)
(a) Crystal structure parameters $a$, $b$ and $z_{\rm Se}$
for $P=0\sim0.6$GPa reported by Millican {\it et al} \cite{MillicanS}.
The Crystal structure parameters at ambient pressure
reported in Ref. \cite{BohmerS} are also shown.
(b) Dispersions on the $k_x$-$k_y$ plane
in the three-dimensional WIEN2k bandstructure for $P=0$ and $P=4$GPa.
The characteristic change in the bandstructure is
reproduced by introducing $\Delta E_{xy}$, 
as shown in Fig. 1 (a) in the main text.
}
\label{fig:AP1}
\end{figure}

To justify the validity of $H_P$,
we perform the band calculation by using the WIEN2k code.
We use the crystal structure of FeSe under pressure 
reported in Ref. \cite{MillicanS} for $0\sim0.6$GPa,
which are shown in Fig. \ref{fig:AP1} (a).
We perform the first principles calculation
using the WIEN2k code for $P=0$GPa and 0.6GPa,
and find that $E_{xy}$-level at $\Gamma$-point
is lifted by 13meV, whereas $E_{xz(yz)}$-level at $\Gamma$-point
is lowered by 9meV by applying 0.6GPa.
That is, the difference $E_{xy}-E_{xz(yz)}$ increases by 25meV
by applying 0.6GPa.
This result is consistent with the theoretical report in
Ref. \cite{Kuroki-zS} that the top of the $d_{xy}$-orbital hole-pocket
is lifted as increasing $z_{\rm Se}$.

%%%%%%%%%%%%%%%%%%%%%%%%%%%%%%%%%
%%%%%%%%%%%%%%%%%%%%%%%%%%%%%%%%%

According to the recent study for FeSe under high pressure
 \cite{MizukamiS},
the lattice constants $a$ and $c$ linearly decrease
in proportion to $P$ for $P<4$GPa.
Here, we perform the band calculations for $P=0$GPa and $4$GPa,
by assuming the linear extrapolations of $a$, $b$ and $z_{\rm Se}$
from Fig. \ref{fig:AP1} (a).
The disperdion on the $k_z=0$ plane
in the three-dimensional WIEN2k bandstructure
are shown in Fig. \ref{fig:AP1} (b).
It is predicted that $E_{xy}-E_{xz(yz)}$ increases by 170meV
by applying 4GPa.
%Such pressure-induced band modification in Fig. \ref{fig:AP1} (b) 
%is qualitatively reproduced in Fig. 1 (a) in the main text
%by introducing $\Delta E_{xy}$.
Similar orbital-dependent change in the bandstructure
%is obtained in the two-dimensional first-principles model
is obtained even if we drop the inter-layer hoppings,
as we show in Fig. \ref{fig:AP5} (b).
%which is introduced in SM: D by dropping the inter-layer hoppings;
Therefore, it is natural to expect that 
the top of the $d_{xy}$-hole pocket,
which is about $40$meV below the Fermi level at 0GPa
\cite{FeSe-ARPES6S},
is shifted to above the Fermi level under pressure.
Thus, the simplified pressure term $H_P$ in the main text is justified.

In the main text, we analyze the total Hamiltonian
$H=H_0+H_P+rH_U$,
where $H_0$ is the kinetic term at $P=0$
introduced in Ref. \cite{Yamakawa-FeSeS},
and $H_U$ is the screened Coulomb interaction due to the valence-bands
obtained by the constraint-RPA (cRPA) method
\cite{MiyakeS}.
We neglect the pressure dependence of $H_U$ 
since the screening would be insensitive to the 
pressure-induced change in the valence-band structure.
Here, we introduce the reduction factor $r$
and the orbital-dependent renormalization factor $z_l$ ($l=1\sim5$).
In the case of $z_l=z$ for all $l$, the Stoner factors for 
the parameters $(r,T)$ at $z=1$ 
are equal to those for $(r/z,zT)$ at $z<1$.
\cite{Yamakawa-FeSeS}.

%add
In this study, we put $z_4 < 1$ and $z_l=1$ ($l\ne4$),
which is consistent with the experimental relation by ARPES.
In the present numerical study at $T\le30$ meV,
we put $1/z_4=1.2$ in order to reproduce the moderate
spin fluctuations at $\q=(\pi,0)$ in FeSe at ambient pressure.
(When $z_4=1$, $\chi^s(\q)$ has the maximum at $\q=(\pi,\pi)$
due to the nesting between eFS1 and eFS2 on $xy$-orbital.)
At $T=50$ meV, $1/z_4$ should be larger than $1.5$ 
to satisfy the relation $\chi^s(\pi,0)>\chi^s(\pi,\pi)$ \cite{Onari-FeSeS}.
The reason is that 
the hFS1,2 shrink at $T=50$ meV due to the temperature-induced shift 
in the chemical potential, 
so $\chi^s_{yz}(\pi,0)$ is relatively reduced. 
Note that the numerical results for $1/z_4=1.2$ at $T= 30$ meV 
are essentially similar to the results for $1/z_4=1.6$ at $T= 50$ meV.

%%%%%%%%%%%%%%%%%%%%%%%%%%%%%%%%%%%%%%%%%%%%%%%%%%%%%%%%
\subsection{B: Full self-consistent vertex correction analysis}

In the main text, 
we analyzed the eight-orbital Hubbard model for FeSe
by using the SC-VC theory.
In this theory, the VC for the spin and charge irreducible
susceptibilities, ${\hat X}^s(q)$ and ${\hat X}^c(q)$,
are calculated self-consistently.
For ${\hat X}^{s,c}(q)$,
both the Maki-Thompson-type VC
and the AL-type VC are analyzed
\cite{Onari-SCVC-S,Yamakawa-FeSeS,Onari-FeSeS}.
%In Fe-based superconductors,
In the main text, 
we drop ${\hat X}^s(q)$ 
and calculate only the charge-channel VC $X_{l,l';m,m'}^c(q)$
for $l,l',m,m'=2\sim4$ except for $l=l'=m=m'=4$,
by following the simplification introduced in Ref. \cite{Onari-SCVC-S}.
This simplification is justified
%for analyzing the orbital fluctuations 
for Fe-based superconductors
as verified in our previous studies
\cite{Yamakawa-FeSeS,Onari-FeSeS},

%%%%%%%%%%%%%%%%%%%%%%%%%%%%%%%%%
\begin{figure}[!htb]
\includegraphics[width=.9\linewidth]{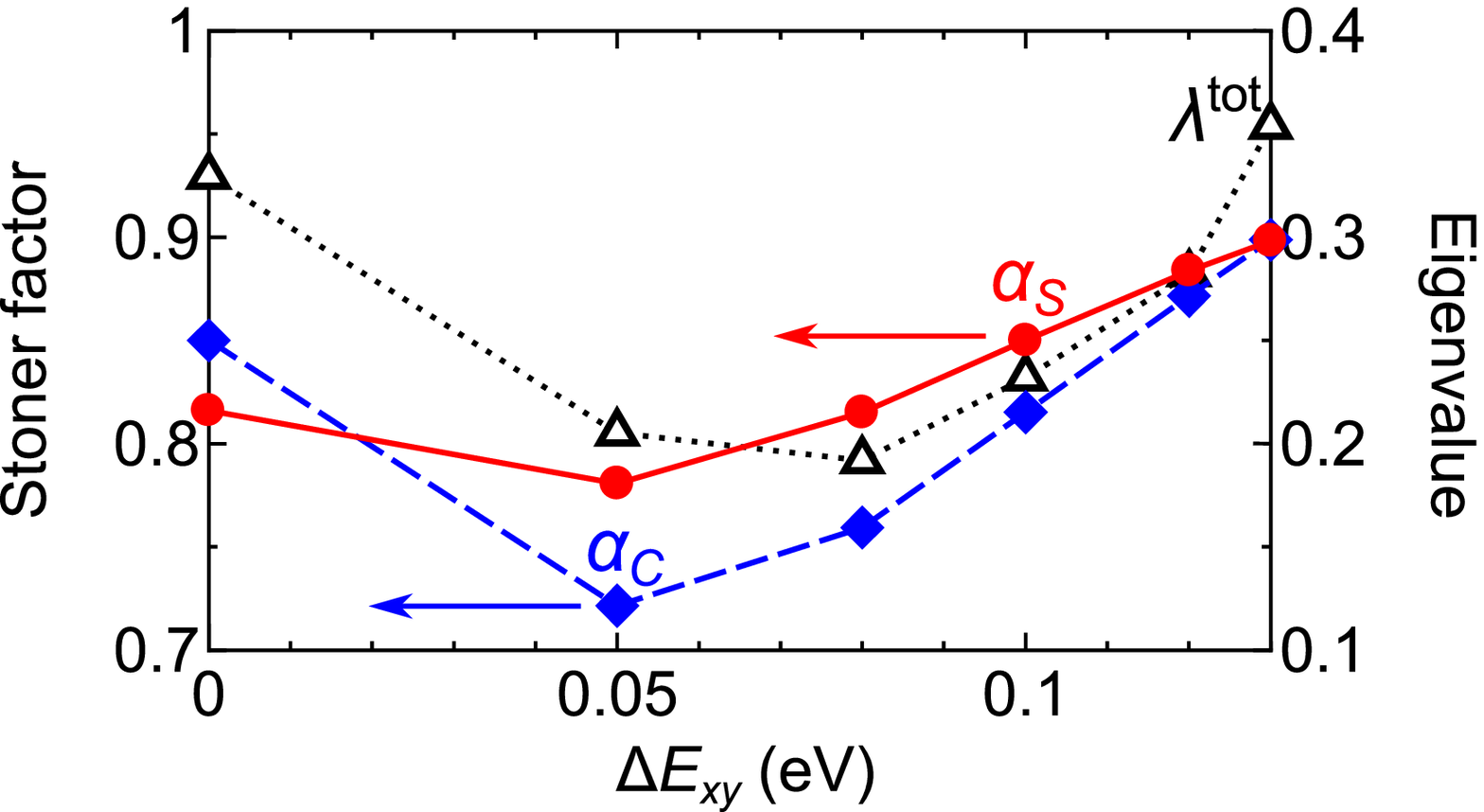}
\caption{
(color online)
Stoner factors ($\a_S$, $\a_C$) and gap equation eigenvalue 
for $V^{\rm tot}$ ($\lambda^{\rm tot}$) as functions of $\Delta E_{xy}$
given by the full SC-VC analysis.
}
\label{fig:AP2}
\end{figure}
%%%%%%%%%%%%%%%%%%%%%%%%%%%%%%%%%

Here, we perform the time-consuming
full self-consistent calculation for 
both ${\hat X}^{s}(q)$ and ${\hat X}^c(q)$ for all $d$-orbitals,
in order to verify the validity of the numerical study in the main text.
To reflect the increase of the bandwidth under pressure,
we put $r=r_0-a\cdot \Delta E_{xy}$, where $r_0=0.287$ and $a=0.19$.
Then, $r=0.287 \ (0.268)$ for $\Delta E_{xy}=0 \ (0.1)$ eV.
In Fig. \ref{fig:AP2},
we show the obtained 
%$\Delta E_{xy}$-dependences of 
Stoner factors $\a_S$ and $\a_C$ for $z_4^{-1}=1.4$ at $T=30$meV. 
For $\Delta E_{xy}=0$, the obtained relation $\a_C>\a_S$
means that the orbital order is realized.
When $\Delta E_{xy}$ increases to 0.05eV, 
both $\a_C$ and $\a_S$ decreases and the relation $\a_S>\a_C$ holds.
For $\Delta E_{xy}>0.05$eV,
both $\a_C$ and $\a_S$ starts to increase, meaning that $T_m$
increases with the pressure, consistently with the phase diagram 
in FeSe under pressure.
In addition, we show the eigenvalue of the $s$-wave
gap equation $\lambda_{\rm tot}$ obtained for $V^{\rm tot}$.
The increment of $\lambda_{\rm tot}$ for $\Delta E_{xy}>0.08$eV
is consistent with the increment of $T_{\rm c}$ under pressure
observed experimentally.
In the present calculation,
the intra-pocket attractive interaction is large whereas
the interaction between hole- and electron-pockets 
is  small because of the cancellation
between attractive and repulsive interactions.
%between $\frac32 I^s>0$ and $-\frac12 I^c+V^{\rm cross}<0$ in $V^{\rm tot}$.
For this reason, we find that both the $s_{++}$-wave and $s_\pm$-wave states 
can appear depending of the model parameters,
and the impurity-induced $s_\pm \rightarrow s_{++}$ crossover
is easily realized.
%Here, with increasing $\Delta E_{xy}$, the $s_{++}$-wave for $\Delta E_{xz}\sim0$
%gradually changes to the $s_{\pm}$-wave state for $\sim0.1$eV.
Note that $\lambda^{\rm tot}$ in Fig. \ref{fig:AP2}
is smaller than unity,
meaning that $\lambda^{\rm tot}$ given in the main text
is overestimated due to the simplified analysis.

%%%%%%%%%%%%%%%%%%%%%%%%%%%%%%%%%%%%%%%%%%%%%%%%%%%%%%%%
\subsection{C: Gap functions for larger orbital polarization}

%%%%%%%%%%%%%%%%%%%%%%%%%%%%%%%%%
\begin{figure}[!htb]
\includegraphics[width=.99\linewidth]{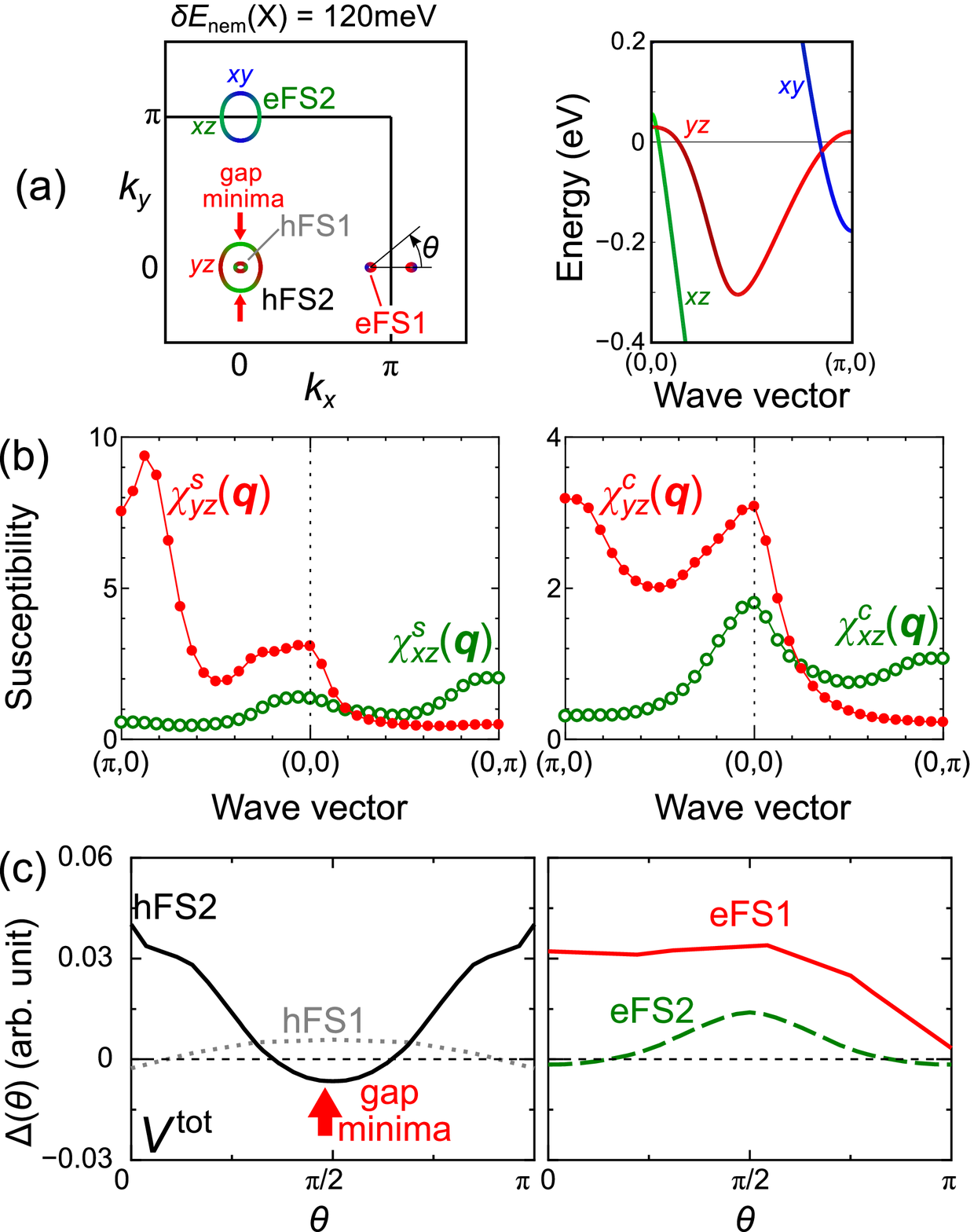}
\caption{
(color online)
(a) FSs and bandstructure for
$\delta E_{\rm nem}(\Gamma)= -30$meV and $\delta E_{\rm nem}({\rm X})= 120$meV.
Here, $\Delta E_{xy}=0$.
(b) $\chi^s_{xz}(\q)$, $\chi^s_{yz}(\q)$, 
$\chi^c_{xz}(\q)$, and $\chi^c_{yz}(\q)$
obtained by the SC-VC theory.
%(c) Gap functions on the FSs obtained for $V^{\Lambda}$
%($\lambda^{\Lambda}=1.44$).
(c) Gap functions obtained for $V^{\rm tot}$
($\lambda^{\rm tot}=0.66$).
}
\label{fig:AP3}
\end{figure}
%%%%%%%%%%%%%%%%%%%%%%%%%%%%%%%%%

%%%%%%%%%%%%%%%%%%%%%%%%%%%%%%%%%
\begin{figure}[!htb]
\includegraphics[width=.99\linewidth]{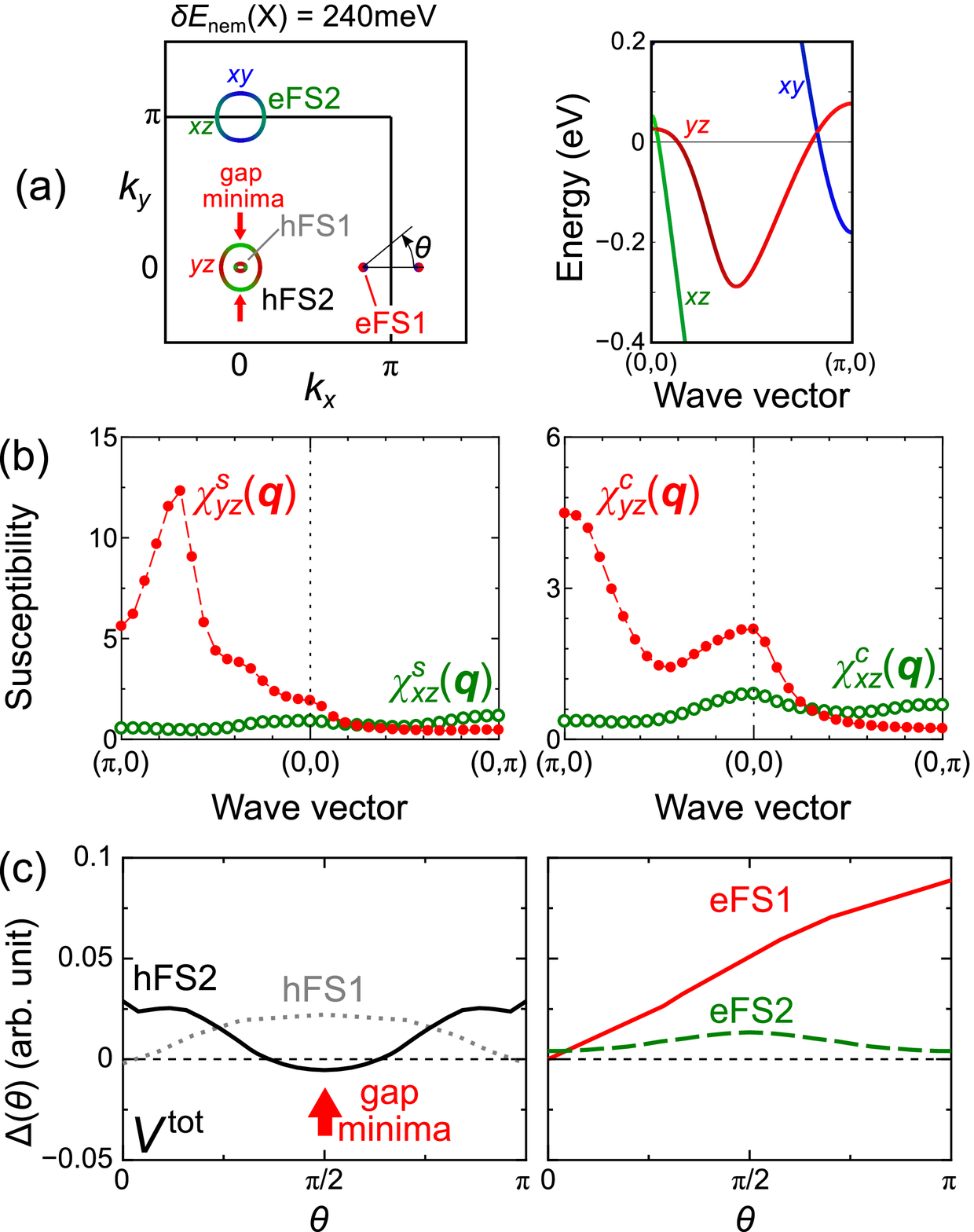}
\caption{
(color online)
(a) FSs and bandstructure for
$\delta E_{\rm nem}(\Gamma)= -30$meV and $\delta E_{\rm nem}({\rm X})= 240$meV.
(b) $\chi^s_{xz}(\q)$, $\chi^s_{yz}(\q)$, 
$\chi^c_{xz}(\q)$, and $\chi^c_{yz}(\q)$
obtained by the SC-VC theory.
%(c) Gap functions on the FSs obtained for $V^{\Lambda}$ 
%($\lambda^{\Lambda}=1.46$).
(c) Gap functions obtained for $V^{\rm tot}$
($\lambda^{\rm tot}=0.79$).
}
\label{fig:AP4}
\end{figure}
%%%%%%%%%%%%%%%%%%%%%%%%%%%%%%%%%

In the main text, we studied the gap function in FeSe 
for $\delta E_{\rm nem}({\rm X})= 60$meV,
in which the horizontally-long electron-like FS
is realized around the X-point.
Here, we study the gap function for $\delta E_{\rm nem}({\rm X})>100$meV,
in which the electron-like FS is divided into two Dirac cones.
We note that the experimentally observed 
orbital splitting is $z\cdot \delta E_{\rm nem}$,
where $z\ (\gtrsim3)$ is the renormalization factor for 
$xz,yz$-orbitals \cite{DMFT1S}.

First, we study the case
$\delta E_{\rm nem}(\Gamma)= -30$meV and $\delta E_{\rm nem}({\rm X})= 120$meV.
The FSs and bandstructure are shown in Fig. \ref{fig:AP3} (a):
The Dirac cones around X-point are electron-like.
The susceptibilities obtained by the SC-VC theory
for $r=0.266$ at $T=30$ meV 
[$(\a_S,\a_C)=(0.953,0.802)$]
are shown in Fig. \ref{fig:AP3} (b).
Here, we dropped the VC for the spin susceptibility 
as we did in the main text.
Figure \ref{fig:AP3} (c) shows the $s$-wave gap function 
for $V^{\rm SC}=V^{\rm tot}$.
Here,
the gap is large on the FSs with large $\dyz$-orbital component,
since large intra-pocket and inter-pocket attractive force are
caused by large $\chi^c_{yz}(\q)$ at $\q\sim{\bm 0}$ and $\q\sim(\pi,0)$.
(We find that $V^{\rm cross}$ give weak repulsive interaction in this case.)
The position of the nodal part on hole-FS is consistent with 
the experimental report \cite{Feng-FeSe-gap-S}.

Next, we study the case
$\delta E_{\rm nem}(\Gamma)= -30$meV and $\delta E_{\rm nem}({\rm X})= 240$meV.
The Dirac cones around X-point in Fig. \ref{fig:AP4} (a) are hole-like.
The susceptibilities for $r=0.266$ at $T=30$ meV 
[$(\a_S,\a_C)=(0.963,0.842)$]
are shown in Fig. \ref{fig:AP4} (b).
Figure \ref{fig:AP4} (c) shows the $s$-wave gap function 
for $V^{\rm SC}=V^{\rm tot}$.
In both cases,
the gap function on the Dirac cone takes the largest value,
which is consistent with the recent experimental report
\cite{BEC-S}.

%%%%%%%%%%%%%%%%%%%%%%%%%%%%%%%%%%%%%%%%%%%%%%%%%%%%%%%%
\subsection{D: Numerical study based on $H_P^{\rm 1st}$ 
given by the first-principles study}

\begin{figure}[!htb]
\includegraphics[width=.99\linewidth]{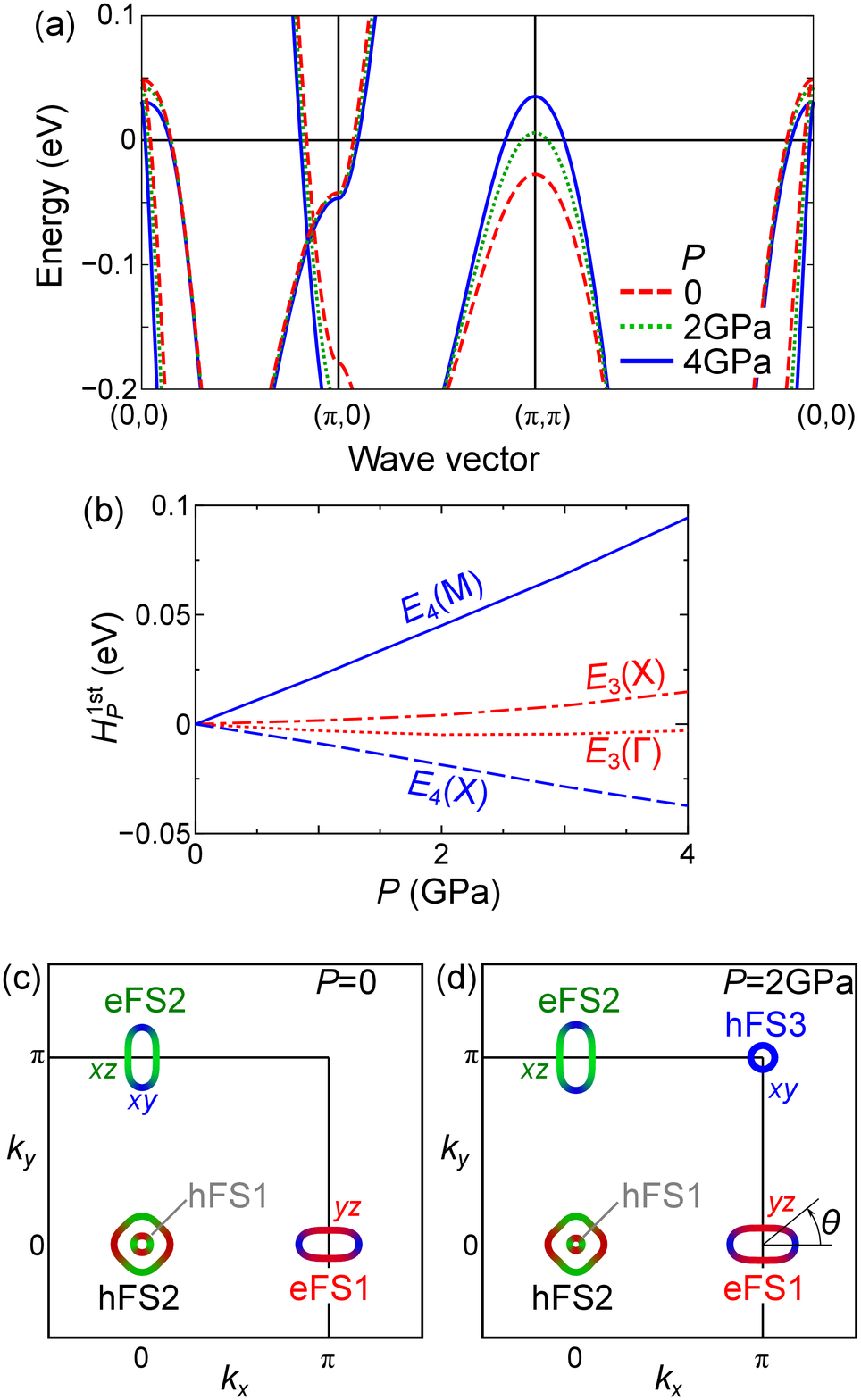}
\caption{
(color online)
(a) Band structure of the two-dimensional FeSe model
$H_0+H_P^{\rm 1st}$ at $P=0$, $2$, and $4$ GPa.
(b) $E_3(\Gamma)$, $E_3({\rm X})$, $E_4({\rm X})$ and $E_4({\rm M})$
given by $H_P^{\rm 1st}(P)$.
(c) FSs at $P=0$ GPa.
(d) FSs at $P=2$ GPa: The hFS3 is the $\dxy$-orbital hole-pocket.
}
\label{fig:AP5}
\end{figure}

Here, we derive the pressure effect Hamiltonian
$H_P^{\rm 1st}\equiv H^{\rm 1st}(P)-H^{\rm 1st}(0)$,
where $H^{\rm 1st}(P)$ is the first principles tight-binding model
for FeSe under $P$ GPa in the tetragonal phase given by the WIEN2k software.
The inter-layer hopping integrals are dropped to obtain the 
two-dimensional model.
Hereafter, we study the total Hamiltonian 
$H=H_0+H^{\rm 1st}(P)+ rH_U$.
Here, we introduce $H_0$ by shifting the orbital level 
$\delta E_l ( \Gamma ), \delta E_l ({\rm X}), \delta E_l ({\rm Y}), 
\delta E_l ({\rm M})$ as
$(-0.24, -0.38, 0.12, 0.5)$ for $l=2$,
$(-0.24, 0.12, -0.38, 0.5)$ for $l=3$, and 
$(0.5, -0.24, -0.24, 0.2)$ for $l=4$ in unit eV.
Here, ${\rm M}=(\pi,\pi)$.
%$(\delta E_{l}(\Gamma),\delta E_{xz}({\rm X}),\delta E_{xz}(\pi,\pi))$
%as $(-0.24,-0.38,0.5)$ for $l=2,3$, and
%$(0.5,-0.24,0.2)$ for $l=4$ 
Figure \ref{fig:AP5} (a) shows the bandstructure
for $P=0$, $2$, and $4$ GPa given by the eigenvalues of
$H_0+H^{\rm 1st}(P)$.
The FSs are sensitively modified by the 
following orbital levels given by  $H^{\rm 1st}(P)$:
$E_2(\Gamma)=E_3(\Gamma)$, $E_2({\rm Y})=E_3({\rm X})$, 
$E_4({\rm X})$, and $E_4({\rm M})$
These $P$-dependences are shown in Fig. \ref{fig:AP5} (b).
Among them, $E_4({\rm M})$ shows the most prominent $P$-dependence.
The obtained FSs for $P=0$ and $2$ GPa are shown in 
Figs. \ref{fig:AP5} (c) and (d), respectively.
Since the top of the $xy$-orbital hole-band at $P=0$ is $-40$ meV
in the present model, the hFS3 appears at $P=2$GPa.

%%%%%%%%%%%%%%%%%%%%%%%%%%%%%%%%%
%%%%%%%%%%%%%%%%%%%%%%%%%%%%%%%%%

\begin{figure}[!htb]
\includegraphics[width=.99\linewidth]{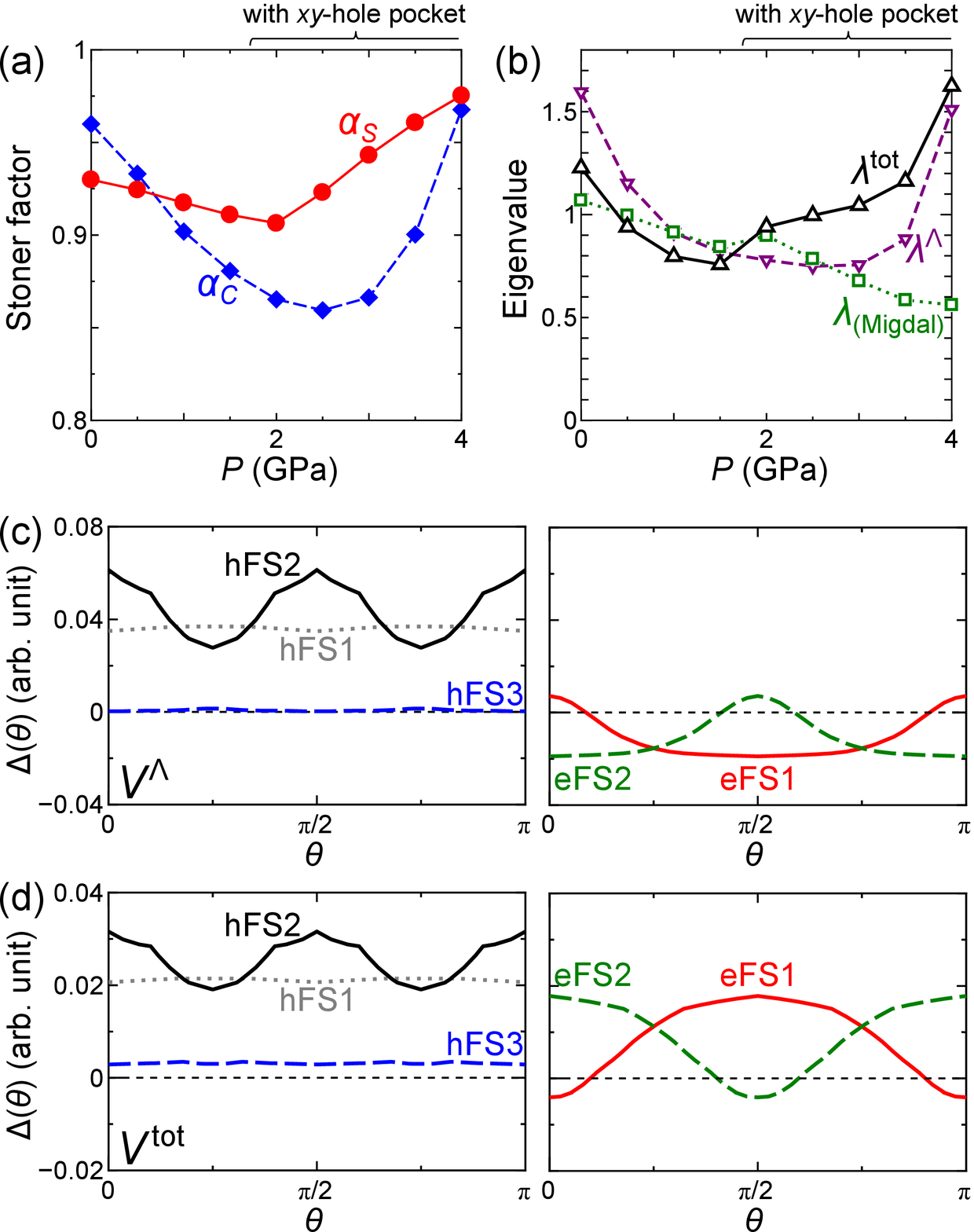}
\caption{
(color online)
(a) Stoner factors $\a_{S,C}$ for $P=0\sim4$ GPa,
at $T=20$meV and $r=0.263$.
(b) Eigenvalues $\lambda^{\Lambda}$ and $\lambda^{\rm tot}$
as functions of $P$.
For comparison, we also show $\lambda_{\rm (Migdal)}$.
(c) $s_\pm$-wave gap given by $V^{\Lambda}$ at $P=4$ GPa.
(d) $s_{++}$-wave gap given by $V^{\rm tot}$ at $P=4$ GPa.
Numerical study is performed in the nonmagnetic state.
}
\label{fig:AP6}
\end{figure}

Figure \ref{fig:AP6} (a) shows the Stoner factors 
$\a_{S,C}$ obtained by the SC-VC theory for $r=0.263$ at $T=20$meV. 
At ambient pressure, $\a_C$ is much larger than $\a_S$.
For $P\lesssim2$GPa, 
$\a_C$ decreases in proportion to $P$,
reflecting the decrease of $X_{yz}^c({\bm 0})\sim \chi^s_{yz}(\Q)$.
In this model, the minimum of $\a_C$ is 0.86 at $P=2.5$ GPa,
which becomes much smaller than that in Fig. 2 (e) in the main text
($\a_C=0.925$ at $\Delta E_{xy}=0.1$eV).
For $P\gtrsim2$GPa, the hFS3 appears,
and then $\a_S$ starts to increase due to the 
nesting between hFS3 and eFS1,2.
In addition, $\a_C$ also starts to increase for $P>2.5$ GPa
since $\chi^s_{xy}(\q)$ also contribute to the ``AL-term on $yz$-orbital
due to $|\Lambda_{3,3;4,4;4,4}^0|^2T\sum_{\q}\{\chi^s_{xy}(\q)\}^2$''.
These results are essentially equivalent to 
Fig. \ref{fig:fig2} (f) in the main text.

Figure \ref{fig:AP6} (b) shows the
eigenvalue of the gap equation obtained for $V^{\rm tot}$ ($\lambda^{\rm tot}$)
and that for $V^{\Lambda}$ ($\lambda^{\Lambda}$).
Both $V^{\rm tot}$ and $V^{\Lambda}$ includes the beyond-Migdal-Eliashberg effects
discussed in the main text.
%the $U$-VC and the AL-type crossing fluctuation term.
For comparison, we also show $\lambda_{\rm (Migdal)}$.
For $P=4$GPa,
the $s_\pm$-wave state with sign reversal between electron- and hole-pockets
is obtained for $V^{\Lambda}$
as shown in Fig. \ref{fig:AP6} (c).
In contrast, the ``$s_{++}$-wave state without sign reversal'' is obtained 
for the total pairing interaction $V^{\rm tot}$
as shown in Fig. \ref{fig:AP6} (d),
since $V^{\rm cross}$ gives large inter-pocket attractive interaction:
%The obtained nodal $s_{++}$-wave state at $P=4$GPa
%is shown in Fig. \ref{fig:AP6} (d).
These results are essentially equivalent to 
Figs. \ref{fig:fig3} (f) and (g) in the main text.

%%%%%%%%%%%%%%%%%%%%%%%%%%%%%%%%%
%%%%%%%%%%%%%%%%%%%%%%%%%%%%%%%%%

The low-energy bandstructure of $H_0$ introduced in this section 
is very similar to that in the main text
near the Fermi level ($|\mu-\e|<0.5$ eV).
(Note that the bandstructure away from the Fermi level
is unknown because of the broadening of the ARPES spectra.)
Even if $H_0$ in the main text are used, 
the obtained $P$-dependences of the Stoner factors 
in Fig. \ref{fig:AP6} (a) are qualitatively unchanged.

%%%%%%%%%%%%%%%%%%%%%%%%%%%%%%%%%%%%%%%%%%%%%%%%%%%%%%%%
\subsection{E: Pairing interaction beyond the Migdal-Eliashberg theory}

In conventional Migdal-Eliashberg (ME) gap equation,
the pairing interaction is given by the 
single-fluctuation exchange process, 
and the VC for the coupling constant ($U$-VC) is dropped.
In the main text, we studied the gap equation 
by including the pairing interaction beyond the ME formalism,
such as the $U$-VC and 
the double-fluctuation exchange processes $V^{\rm cross}(\k,\p)$.
Here, we explain the analytic expressions for these beyond ME processes
\cite{Yamakawa-eFeSeS}.

The analytic expression of $V^{\Lambda}$ in Fig. 3 (b) in the main text is
\begin{eqnarray}
{\hat V}^{\Lambda}(k,p)
=\frac32{\hat I}^{\Lambda,s}(k,p)
-\frac12{\hat I}^{\Lambda,c}(k,p)-{\hat U}^{0s},
\end{eqnarray}
where 
\begin{eqnarray}
{\hat I}^{\Lambda,x}(k,p)= {\hat \Lambda}^x(k,p)
{\hat I}^{x}(k-p){\hat {\bar \Lambda}}^x(-k,-p).
\end{eqnarray}
Here, $\Lambda^x$ ($x=s,c$) is the $U$-VC shown in Fig. 3 (d).
From now on, we explain the analytic expressions for the
$U$-VC due to the Maki-Thompson (MT) and Aslamazov-Larkin (AL) processes
which were already given in Ref. \cite{Yamakawa-eFeSeS}.
First, we explain the  charge- and spin-channel MT-terms:
\begin{eqnarray}
	\Lambda^{{\rm MT}, c}_{l,l';m,m'} (k,k')
  	&=& \frac{T}{2} \sum_{p} \sum_{a,b}
	\left\{
		I^{c}_{b,l';a,l} (p) + 3 I^{s}_{b,l';a,l} (p)
	\right\}
\nonumber \\
	&&\times G_{a,m} (k+p) G_{m',b} (k'+p) 
\label{eqn:UMTc} ,
\end{eqnarray}

\begin{eqnarray}
	\Lambda^{{\rm MT}, s}_{l,l';m,m'} (k,k')
  	&=& \frac{T}{2} \sum_{p} \sum_{a,b}
	\left\{
		I^{c}_{b,l';a,l} (p) - I^{s}_{b,l';a,l} (p)
	\right\}
\nonumber \\
	&&\times G_{a,m} (k+p) G_{m',b} (k'+p) 
\label{eqn:UMTs} ,
\end{eqnarray}
where
${\hat I}^x(q)= {\hat U}^{0x}{\hat \chi}^x(q){\hat U}^{0x}+{\hat U}^{0x}$.

Next, we explain the charge- and spin-channel AL-terms:
\begin{eqnarray}
	&&\Lambda^{{\rm AL}, c}_{l,l';m,m'} (k,k')
\nonumber \\
&&\quad = \frac{T}{2} \sum_{p} \sum_{a,b,c,d,e,f}
	G_{a,b} (k'-p) {\Lambda^0}'_{m,m';c,d;e,f} (k-k',p)
\nonumber \\
	&&\quad \times
	\left\{
	    I^{c}_{l,a;c,d} (k-k'+p) I^{c}_{b,l';e,f} (-p) \right.
\nonumber \\
	&&\quad \left. + 3 I^{s}_{l,a;c,d} (k-k'+p) I^{s}_{b,l';e,f} (-p)
	\right\}
\label{eqn:UALc} ,
%\Lambda^{{\rm MT},c}_{l,l';m,m'} (k,k')
%&=& \frac{T}{2} \sum_{q} \sum_{a,b} 
%\big\{ I^{c}_{a,l;b,l'}(q)+3 I^{s}_{a,l;b,l'}(q) \big\} 
%\nonumber \\
%& &\times G_{a,m}(k+q)G_{m',b} (k'+q),
% \\
%\Lambda^{{\rm MT},s}_{l,l';m,m'} (k, k')
%&=& \frac{T}{2} \sum_{q} \sum_{a,b} 
%\big\{ I^{c}_{a,l;b,l'}(q) - I^{s}_{a,l;b,l'}(q) \big\} 
%\nonumber \\
%& &\times G_{a,m'} (k+q) G_{m,b} (k'+q),
\end{eqnarray}
\begin{eqnarray}
	&&\Lambda^{{\rm AL}, s}_{l,l';m,m'} (k,k')
\nonumber \\
	&&\quad = \frac{T}{2} \sum_{p} \sum_{a,b,c,d,e,f} G_{a,b} (k'-p) 
	{\Lambda^0}'_{m,m';c,d;e,f} (k-k',p)
\nonumber \\
	&& \quad \times
	\left\{
	  I^{c}_{l,a;c,d} (k-k'+p) I^{s}_{b,l';e,f} (-p)
	\right.	
\nonumber \\
	&& \quad \left.
	+ I^{s}_{l,a;c,d} (k-k'+p) I^{c}_{b,l';e,f} (-p)
	\right\}
\nonumber \\
	&& \quad
+ \delta\Lambda^{{\rm AL}, s}_{l,l';m,m'} (k,k')
%	\left.
%	+ 2 I^{s}_{l,a;c,d} (k-k'+p) I^{s}_{b,l';e,f} (-p)
%	\Lambda''_{m,m';c,d;e,f} (k-k';p)
%	\right] 
%\nonumber \\
%
%&&\Lambda^{{\rm AL},c}_{l,l';m,m'} (k, k')
%=\frac{T}{2} \sum_{q} \sum_{abcdef} \nonumber \\
%&& \times \big\{ \Lambda_{abcdef} (k - k', q) + \Lambda_{fcbeda} (k - k', - q - k + k') \big\} \nonumber \\
%&& \times \big\{ I^{c}_{bclg} (q + k - k') I^{c}_{mhed} (q) + 3 I^{s}_{bclg} (q + k - k') I^{s}_{mhed} (q) \big\} \nonumber \\
%&& \times G_{gh} (k' - q) ,
%\label{eqn:UALc} \\
%&&\Lambda^{{\rm AL},s}_{l'm'lm} (k, k')
%= \frac{T}{2} \sum_{q} \sum_{abcdef} \nonumber \\
%&& \times \big\{ \Lambda_{abcdef} (k - k', q) + \Lambda_{fcbeda} (k - k', - q - k + k') \big\} \nonumber \\
%&& \times \big\{ I^{s}_{bclg} (q + k - k') I^{c}_{mhed} (q) + I^{c}_{bclg} (q + k - k') )I^{s}_{mhed} (q) \big\} \nonumber \\
%&& \times G_{gh} (k' - q)
%\nonumber \\
%&& + \delta \Lambda^{{\rm AL},s}_{l,l';m,m'} (k, k'),
\label{eqn:UALs} ,
\end{eqnarray}
where 
${\Lambda^0}'_{m,m';c,d;g,h}(q,p)\equiv
\Lambda^0_{c,h;m,g;d,m'}(q,p)+\Lambda^0_{g,d;m,c;h,m'}(q,-p-q)$.
${\hat \Lambda}^0$ is the three-point vertex in Fig. 2 (b) in the main text.
% introduced in the SI:A.
The last term in Eq. (\ref{eqn:UALs}) is given as
\begin{eqnarray}
	&&\delta \Lambda^{{\rm AL}, s}_{l,l';m,m'} (k,k')
	= T \sum_{p} \sum_{a,b,c,d,e,f} G_{a,b} (k'-p) 
 \nonumber \\
	&&\quad \times I^{s}_{l,a;c,d} (k-k'+p) I^{s}_{b,l';e,f} (-p)
	{\Lambda^0}''_{m,m';c,d;e,f} (k-k',p) ,
\nonumber \\
\end{eqnarray}
which is found to be very small 
\cite{Onari-SCVC-S}. 
Here, ${\Lambda^0}''_{m,m';c,d;g,h}(q,p)\equiv
\Lambda^0_{c,h;m,g;d,m'}(q,p)-\Lambda^0_{g,d;m,c;h,m'}(q,-p-q)$.
%$\delta \Lambda^{{\rm AL},s}_{l'm'lm} (k, k')
%=\frac{T}{2} \sum_{q} \sum_{abcdef} 
%\big\{ \Lambda_{abcdef} (k - k', q) - 
%\Lambda_{fcbeda} (k - k', - q - k + k') \big\} 
% 2 I^{s}_{bclg} (q + k - k') I^{s}_{mhed} (q) G_{gh} (k' - q)$,

%\cite{Onari-SCVC-S}.

The ($U^0$)-linear terms in Eqs. (\ref{eqn:UMTc}) and (\ref{eqn:UMTs})
should be dropped to avoid the double counting of the RPA-type diagrams.
We also carefully drop the double counting $(U^0)^2$-terms
included in both MT and AL terms.
We verified numerically that the large charge-channel $U$-VC
($|\Lambda^c|^2\gg1$) originates from the $\chi^s$-square term 
in Eq. (\ref{eqn:UALc}).
We also verified that the 
relation $|\Lambda^s|^2\ll1$ is realized mainly by the $(U^0)^2$-term
\cite{Yamakawa-eFeSeS}.

%%%%%%%%%%
Next, we explain the double-fluctuation exchange process
$V^{\rm cross}(k,p)$, which gives sizable pairing interaction
in Fe-based superconductors \cite{Yamakawa-eFeSeS}.
It is schematically expressed as Fig. 3 (c) in the main text.
Physically, $V^{\rm cross}(k,p)$ represents the 
pairing interaction due to the ``multi-fluctuation exchange processes''.
This term is expressed as \cite{Yamakawa-eFeSeS}
\begin{eqnarray}
	&&V^{\rm cross}_{l,l';m,m'} ( k, p )
	= \frac{T}{4} \sum_{q} \sum_{a,b,c,d}
	G_{a,b} (p-q)
	G_{c,d} (-k-q)
\nonumber \\
	&& \times
	\left\{
        3 I'^{s}_{l,a;m,d} (k-p+q) I'^{s}_{b,l';c,m'} (-q) 
        \right.
\nonumber \\
	&&\quad 
	+ 3 I'^{s}_{l,a;m,d} (k-p+q) I'^{c}_{b,l';c,m'} (-q) 
\nonumber \\
	&&\quad  
        + 3 I'^{c}_{l,a;m,d} (k-p+q) I'^{s}_{b,l';c,m'} (-q) 
\nonumber \\
	&&\quad \left.
	-   I'^{c}_{l,a;m,d} (k-p+q) I'^{c}_{b,l';c,m'} (-q) 
	\right\} 
\label{eqn:Vcross} ,
\end{eqnarray}
where we put ${\hat I}'^x= {\hat I}^x-{\hat U}^{0x}$
to avoid the double counting of diagrams included in other terms.
%Here, the $U^2$-term of $V^{\rm cross}$ should be dropped
%since it is already included in the interaction in Fig. \ref{fig:fig1} (b).
According to Eq. (\ref{eqn:Vcross}),
$V^{\rm cross}_{m,m;m,m}$ can take large negative value when 
the spin fluctuations develop on the $m$-orbital
\cite{Yamakawa-eFeSeS}.

%%%%%%%%%%%%%%%%%%%%%%%%
%references
%%%%%%%%%%%%%%%%%%%%%%%%

\end{document}